\documentclass[sigconf,authorversion]{acmart}

\usepackage[all]{nowidow}
\usepackage{xspace}

\usepackage{bm}

\usepackage[acronym]{glossaries-extra}
\setabbreviationstyle[acronym]{long-short}

\newglossaryentry{computer}
{
  name=computer,
  description={is a programmable machine that receives input,
               stores and manipulates data, and provides
               output in a useful format}
}

\newacronym[\glslongpluralkey={Systems on a Chip}]{soc}{SoC}{System on a Chip}
\newacronym{alu}{ALU}{Arithmetic Logic Unit}
\newacronym{btc}{BTC}{Branch Target Cache}
\newacronym{cisc}{CISC}{Complex Instruction Set Computing}
\newacronym{cpsr}{CPSR}{Current Program Status Register}
\newacronym{crtm}{CRTM}{Core Root of Trust for Measurement}
\newacronym{dma}{DMA}{Direct Memory Access}
\newacronym{dsb}{DSB}{Data Synchronization Barrier}
\newacronym{dsp}{DSP}{Digital Signal Processor}
\newacronym{dtlb}{DTLB}{Data Translation Lookaside Buffer}
\newacronym{isb}{ISB}{Instruction Synchronization Barrier}
\newacronym{itlb}{ITLB}{Instruction Translation Lookaside Buffer}
\newacronym{jtag}{JTAG}{Joint Test Action Group}
\newacronym{lr}{LR}{Link Register}
\newacronym{mitm}{MitM}{Man-in-the-Middle}
\newacronym{mmc}{MMC}{MultiMediaCard}
\newacronym{mmu}{MMU}{Memory Management Unit}
\newacronym{mpu}{MPU}{Memory Protection Unit}
\newacronym{msb}{MSB}{Most Significant Bit}
\newacronym{nop}{NOP}{No Operation}
\newacronym{pc}{PC}{Program Counter}
\newacronym{pin}{PIN}{Personal Identification Number}
\newacronym{poco}{PoC}{Point of Coherency}
\newacronym{pou}{PoU}{Point of Unification}
\newacronym{risc}{RISC}{Reduced Instruction Set Computing}
\newacronym{rom}{ROM}{Read-Only Memory}
\newacronym{sdio}{SDIO}{Secure Digital Input Output}
\newacronym{sd}{SD}{Secure Digital}
\newacronym{simd}{SIMD}{Single Instruction Multiple Data}
\newacronym{spi}{SPI}{Serial Peripheral Interface}
\newacronym{spsr}{SPSR}{Saved Program Status Register}
\newacronym{sp}{SP}{Stack Pointer}
\newacronym{sram}{SRAM}{Static Random-access Memory}
\newacronym{sr}{SR}{Status Register}
\newacronym{tcg}{TCG}{Trusted Computing Group}
\newacronym{tcp/ip}{TCP/IP}{Transmission Control Protocol/Internet Protocol}
\newacronym{tlb}{TLB}{Translation Lookaside Buffer}
\newacronym{tpm}{TPM}{Trusted Platform Module}
\newacronym{ftpm}{fTPM}{Firmware Trusted Platform Module}
\newacronym[\glslongpluralkey={Virtual Addresses}]{va}{VA}{Virtual Address}
\newacronym{wxorx}{W$\string^$X}{Write XOR Execute}
\newacronym{aes}{AES}{Advanced Encryption Standard}
\newacronym{amp}{AMP}{Asymmetric Multi-Processing}
\newacronym{apdu}{APDU}{Application Protocol Data Unit}
\newacronym{api}{API}{Application Programming Interface}
\newacronym{ap}{AP}{Application Processor}
\newacronym{asid}{ASID}{Address Space Identifier}
\newacronym{vmid}{VMID}{Virtual Machine Identifier}
\newacronym{aslr}{ASLR}{Address Space Layout Randomization}
\newacronym{bsd}{BSD}{Berkeley Software Distribution}
\newacronym{bts}{BTS}{Branch Trace Store}
\newacronym[\glslongpluralkey={Certificate Authorities}]{ca}{CA}{Certificate Authority}
\newacronym{cbc}{CBC}{Cipher Block Chaining}
\newacronym{cfb}{CFB}{Cipher Feedback}
\newacronym{cff}{CFF}{Compact Font Format}
\newacronym{cfi}{CFI}{Control Flow Integrity}
\newacronym{cpi}{CPI}{Code Pointer Integrity}
\newacronym{cf}{CF}{Control Flow}
\newacronym{cpu}{CPU}{Central Processing Unit}
\newacronym{ctr}{CTR}{Counter}
\newacronym{dac}{DAC}{Discretionary Access Control}
\newacronym{dfu}{DFU}{Device Firmware Upgrade}
\newacronym{drm}{DRM}{Digital Rights Management}
\newacronym{eac}{EAC}{Extended Access Control}
\newacronym{ecb}{ECB}{Electronic Code Book}
\newacronym{ecc}{ECC}{Elliptic Curve Cryptography}
\newacronym{essiv}{ESSIV}{Encrypted Salt-Sector Initialization Vector}
\newacronym{fde}{FDE}{Full Disk Encryption}
\newacronym{fiq}{FIQ}{Fast Interrupt}
\newacronym{ftl}{FTL}{Flash Translation Layer}
\newacronym{gcm}{GCM}{Galois/Counter Mode}
\newacronym{gid}{GID}{Device Group ID}
\newacronym{gpu}{GPU}{Graphics Processing Unit}
\newacronym{gui}{GUI}{Graphical User Interface}
\newacronym{hci}{HCI}{Host Controller Interface}
\newacronym{hfs}{HFS}{Hierarchical File System}
\newacronym{hsm}{HSM}{Hardware Security Module}
\newacronym{http}{HTTP}{Hypertext Transfer Protocol}
\newacronym{hvc}{HVC}{Hypervisor Call}
\newacronym{i2c}{I$^{2}$C}{Inter-Integrated Circuit}
\newacronym{icc}{ICC}{Inter Component Communication}
\newacronym[\glslongpluralkey={International Mobile Equipment Identities}]{imei}{IMEI}{International Mobile Equipment Identity}
\newacronym[\glslongpluralkey={Intermediate Physical Addresses}]{ipa}{IPA}{Intermediate Physical Address}
\newacronym{ipc}{IPC}{Inter-Process Communication}
\newacronym{irq}{IRQ}{Interrupt Request}
\newacronym{iv}{IV}{Initialization Vector}
\newacronym{jit}{JIT}{Just-In-Time}
\newacronym{jni}{JNI}{Java Native Interface}
\newacronym{lbr}{LBR}{Last Branch Record}
\newacronym{led}{LED}{Light-Emitting Diode}
\newacronym{lfsr}{LFSR}{Linear Feedback Shift Register}
\newacronym{loc}{LOC}{Lines of Code}
\newacronym{lpae}{LPAE}{Large Physical Address Extension}
\newacronym{lpddr2}{LPDDR2 SDRAM}{Low-Power Double Data Rate Synchronous Dynamic Random Access Memory}
\newacronym{lru}{LRU}{Least Recently Used}
\newacronym{maac}{MAC}{Mandatory Access Control}
\newacronym{mac}{MAC}{Media Access Control}
\newacronym{mcu}{MCU}{Micro Controller Unit}
\newacronym{mno}{MNO}{Mobile Network Operator}
\newacronym{nfc-wi}{NFC-WI}{Near Field Communication Wired Interface}
\newacronym{nfc}{NFC}{Near Field Communication}
\newacronym{nic}{NIC}{Network Interface Controller}
\newacronym{nstid}{NSTID}{Non-Secure Table Identifier}
\newacronym{ns}{NS}{Non-Secure}
\newacronym{nx}{NX}{No eXecute}
\newacronym{ocram}{OCRAM}{On-Chip RAM}
\newacronym{ofb}{OFB}{Output Feedback}
\newacronym{os}{OS}{Operating System}
\newacronym{ota}{OTA}{Over-The-Air}
\newacronym{otp}{OTP}{One-Time Password}
\newacronym{pa}{PA}{Pointer Authentication}
\newacronym{pake}{PAKE}{Password-Authenticated Key Agreement}
\newacronym{pbkdf2}{PBKDF2}{Password-Based Key Derivation Function 2}
\newacronym{pcsc}{PC/SC}{Personal Computer/Smart Card}
\newacronym{pic}{PIC}{Position-Independent Code}
\newacronym{pid}{PID}{Process Identifier}
\newacronym{pki}{PKI}{Public Key Infrastructure}
\newacronym{pl}{PL}{Privilege Level}
\newacronym{el}{EL}{Exception Level}
\newacronym{poc}{PoC}{Proof of Concept}
\newacronym{prng}{PRNG}{Pseudo-Random Number Generator}
\newacronym{qes}{QES}{Qualified Electronic Signature}
\newacronym{ram}{RAM}{Random Access Memory}
\newacronym{dram}{DRAM}{Dynamic Random Access Memory}
\newacronym{rng}{RNG}{Random Number Generator}
\newacronym{rop}{ROP}{Return-Oriented Programming}
\newacronym{scr}{SCR}{Secure Configuration Register}
\newacronym{scu}{SCU}{Snoop Control Unit}
\newacronym{sdk}{SDK}{Software Development Kit}
\newacronym{se}{SE}{Secure Element}
\newacronym{sicc}{SICC}{Secure Interoperable Chip Card Terminal}
\newacronym{sim}{SIM}{Subscriber Identification Module}
\newacronym{slat}{SLAT}{Second Level Address Translation}
\newacronym{smc}{SMC}{Secure Memory Card}
\newacronym{smp}{SMP}{Symmetric Multiprocessing}
\newacronym{sso}{SSO}{Single Sign On}
\newacronym{swp}{SWP}{Single Wire Protocol}
\newacronym{tan}{TAN}{Transaction Authentication Number}
\newacronym{ta}{TA}{Terminal Authentication}
\newacronym{tcb}{TCB}{Trusted Computing Base}
\newacronym{tee}{TEE}{Trusted Execution Environment}
\newacronym{ree}{REE}{Rich Execution Environment}
\newacronym{tlc}{TLC}{Trustlet Connector}
\newacronym{tls}{TLS}{Transport Layer Security}
\newacronym{trng}{TRNG}{True Random Number Generator}
\newacronym{udid}{UDID}{Unique Device Identifier}
\newacronym{uicc}{UICC}{Universal Integrated Circuit Card}
\newacronym{uid}{UID}{Device Unique ID}
\newacronym{ui}{UI}{User Interface}
\newacronym{urb}{URB}{USB Request Block}
\newacronym{usb}{USB}{Universal Serial Bus}
\newacronym{uuid}{UUID}{Universally Unique Identifier}
\newacronym{vmi}{VMI}{Virtual Machine Introspection}
\newacronym{vmm}{VMM}{Virtual Machine Monitor}
\newacronym{vm}{VM}{Virtual Machine}
\newacronym{vpn}{VPN}{Virtual Private Network}
\newacronym{xn}{XN}{Execute Never}
\newacronym{pxn}{PXN}{Privileged Execute Never}
\newacronym{uxn}{UXN}{Unprivileged Execute Never}
\newacronym{smmu}{SMMU}{System Memory Management Unit}
\newacronym{dep}{DEP}{Data Execution Prevention}
\newacronym{cra}{CRA}{Code-Reuse Attack}
\newacronym{dop}{DOP}{Data-Oriented Programming}
\newacronym{iommu}{IOMMU}{Input-Output Memory Management Unit}
\newacronym{crc}{CRC}{Cyclic Redundancy Check}
\newacronym{sev}{SEV}{Secure Encrypted Virtualization}
\newacronym{sev-es}{SEV-ES}{SEV Encrypted State}
\newacronym{vmcb}{VMCB}{Virtual Machine Control Block}
\newacronym{rce}{RCE}{Remote Code Execution}
\newacronym{iram}{iRAM}{Internal RAM}
\newacronym{sme}{SME}{Secure Memory Encryption}
\newacronym{tsme}{TSME}{Transparent SME}
\newacronym{sgx}{SGX}{Software Guard Extensions}
\newacronym{psp}{AMD-SP}{AMD Secure Processor}
\newacronym[\glslongpluralkey={Guest-Physical Addresses}]{gpa}{GPA}{Guest-Physical Address}
\newacronym[\glslongpluralkey={Host-Physical Addresses}]{hpa}{HPA}{Host-Physical Address}
\newacronym{kvm}{KVM}{Kernel-based Virtual Machine}
\newacronym[\glslongpluralkey={Page Table Entries}]{pte}{PTE}{Page Table Entry}
\newacronym[\glslongpluralkey={Identities}]{id}{ID}{Identity}
\newacronym{eap}{EAP}{Extensible Authentication Protocol}
\newacronym{wlan}{WLAN}{Wireless Local Area Network}
\newacronym{isa}{ISA}{Instruction Set Architecture}
\newacronym{sctlr}{SCTLR}{System Control Register}
\newacronym{jvm}{JVM}{Java Virtual Machine}
\newacronym{uefi}{UEFI}{Unified Extensible Firmware Interface}
\newacronym{io}{I/O}{Input/Output}
\newacronym{mee}{MEE}{Memory Encryption Engine}
\newacronym{prm}{PRM}{Processor Reserved Memory}
\newacronym{epc}{EPC}{Enclave Page Cache}
\newacronym{epcm}{EPCM}{Enclave Page Cache Map}
\newacronym{secs}{SECS}{SGX Enclave Control Structure}
\newacronym{tcs}{TCS}{Thread Control Structure}
\newacronym{smm}{SMM}{Secure Management Mode}
\newacronym{tzma}{TZMA}{TrustZone Memory Adapter}
\newacronym{tzasc}{TZASC}{TrustZone Address Space Controller}
\newacronym{iot}{IoT}{Internet of Things}
\newacronym{gps}{GPS}{Global Positioning System}
\newacronym{cve}{CVE}{Common Vulnerabilities and Exposures}
\newacronym{wpa2}{WPA2}{Wi-Fi Protected Access 2}
\newacronym{cet}{CET}{Control-flow Enforcement Technology}
\newacronym{mpx}{MPX}{Memory Protection Extensions}
\newacronym{tme}{TME}{Total Memory Encryption}
\newacronym{ssd}{SSD}{Solid-State Drive}
\newacronym{cfg}{CFG}{Control Flow Graph}
\newacronym{aead}{AEAD}{Authenticated Encryption with Associated Data}

\newacronym{tcms}{TCMS}{Timestamp-based Complete Memory Safety}
\newacronym{pac}{PAC}{Pointer Authentication Code}

\newacronym{mesh}{MESH}{Memory-Efficient Safe Heap}
\newacronym{ir}{IR}{Intermediate Representation}
\newacronym{asan}{ASan}{AddressSanitizer}
\newacronym{hwasan}{HWASan}{Hardware-assisted AddressSanitizer}
\newacronym{cwe}{CWE}{Common Weakness Enumeration}
\newacronym{lto}{LTO}{Link Time Optimization}
\newacronym{got}{GOT}{Global Offset Table}
\newacronym{gppt}{GCPT}{Global CryptSan-Pointer Table}
\newacronym{tbi}{TBI}{Top Byte Ignore}
\newacronym{mte}{MTE}{Memory Tagging Extension}

\glsdisablehyper %

\usepackage{mathtools}

\usepackage[shortlabels]{enumitem}

\usepackage{listings}

\microtypecontext{spacing=nonfrench}

\usepackage{pifont}

\newcommand*{\eg}{e.g.,\@\xspace}
\newcommand*{\ie}{i.e.,\@\xspace}
\newcommand{\Cpp}{C\texttt{++}\xspace}

\newcommand{\Name}{CryptSan\xspace}

\newcommand{\Title}{\Name: Leveraging ARM Pointer Authentication for Memory Safety in C/C\texttt{++}}
\hypersetup{                    %
  pdftitle={\Title},
  pdfauthor={Konrad Hohentanner, Philipp Zieris, Julian Horsch},
}

\copyrightyear{2023}
\acmYear{2023}
\setcopyright{rightsretained}
\acmConference[SAC '23]{The 38th ACM/SIGAPP Symposium on Applied Computing}{March 27-March 31, 2023}{Tallinn, Estonia}
\acmBooktitle{The 38th ACM/SIGAPP Symposium on Applied Computing (SAC '23), March 27-April 2, 2023, Tallinn, Estonia}\acmDOI{10.1145/3555776.3577635}
\acmISBN{978-1-4503-9517-5/23/03}

\begin{document}

\title{\Title}

\author{Konrad Hohentanner}
\email{konrad.hohentanner@aisec.fraunhofer.de}
\orcid{0000-0003-2283-6071}
\affiliation{%
	\institution{Fraunhofer AISEC}
	\streetaddress{}
	\city{Garching, near Munich}
	\state{}
	\postcode{}
	\country{Germany}
}
\author{Philipp Zieris}
\email{philipp.zieris@aisec.fraunhofer.de}
\orcid{0000-0001-9658-1572}
\affiliation{%
	\institution{Fraunhofer AISEC}
	\streetaddress{}
	\city{Garching, near Munich}
	\state{}
	\postcode{}
	\country{Germany}
}
\author{Julian Horsch}
\email{julian.horsch@aisec.fraunhofer.de}
\orcid{0000-0001-9018-7048}
\affiliation{%
	\institution{Fraunhofer AISEC}
	\streetaddress{}
	\city{Garching, near Munich}
	\state{}
	\postcode{}
	\country{Germany}
}

\begin{abstract}
Memory safety bugs remain in the top ranks of security vulnerabilities, even
after decades of research on their detection and prevention. Various
mitigations have been proposed for C/\Cpp, ranging from language dialects to
instrumentation. Among these, compiler-based instrumentation is particularly
promising, not requiring manual code modifications and being able to achieve
precise memory safety. Unfortunately, existing compiler-based solutions
compromise in many areas, including performance but also usability and memory
safety guarantees. New developments in hardware can help improve performance
and security of compiler-based memory safety. ARM Pointer Authentication,
added in the ARMv8.3 architecture, is intended to enable hardware-assisted
\gls{cfi}. But since its operations are generic, it also enables
other, more comprehensive hardware-supported runtime integrity approaches. As
such, we propose \Name, a memory safety approach based on ARM Pointer
Authentication. \Name uses pointer signatures to retrofit memory safety
to C/\Cpp programs, protecting heap, stack, and globals against temporal and
spatial vulnerabilities. We present a full LLVM-based prototype
implementation, running on an M1 MacBook Pro, \ie on actual ARMv8.3 hardware.
Our prototype evaluation shows that the system outperforms similar approaches
under real-world conditions. This, together with its interoperability with
uninstrumented libraries and cryptographic protection against attacks on
metadata, makes \Name a viable solution for retrofitting memory safety to
C/\Cpp programs.
\glsreset{cfi}
\end{abstract}




\maketitle

\newcommand{\hwasan}{HWASan\xspace}

\section{Introduction}
Despite presenting an old and well-known problem, vulnerabilities caused by unsafe
programming languages are still dominant. In memory-unsafe languages, such as C and \Cpp,
programming errors can lead to
various memory corruptions and, in many cases, to dangerous exploits.
MITRE lists out-of-bound writes as number one weakness in its 2022 \gls{cwe}
top list \cite{CWE}, followed by out-of-bound reads and use-after-free on
ranks 5 and 7, respectively.
In recent years, safe language alternatives, such as Rust and Swift,
emerged, but C and \Cpp are widely used and hard to replace,
especially in the embedded domain and for low-level components including
\gls{os} kernels.

Memory corruptions can be divided into spatial and temporal issues.
In a \emph{spatial} memory corruption, a pointer is used to read or write outside
the bounds of its pointed-to object, for example, in a \emph{buffer
	overflow}. In a \emph{temporal} memory corruption, a pointer is used
to access an object that has already been freed, for example, in a
\emph{use-after-free} attack.
Such memory corruptions are typically the first step in more elaborate exploits, where an
attacker tries to corrupt a data or code pointer, to leak 
information or to divert the program's control flow.

In the past decades, numerous approaches have been proposed to mitigate the
effects of memory corruption exploits. Some of the resulting protections, such
as \glsdesc{dep} and \glsdesc{aslr}~\cite{LHBF14}, have proven to provide reasonable trade-offs, and
hence find wide adoption in current systems. These approaches harden
applications by forcing the attacker to reuse existing code and guess 
code locations, without imposing compatibility issues or performance
penalties. Other solutions try to enforce \gls{cfi} using various policies
and techniques to detect deviations in indirect control flow transfers~\cite{BCN+17,BZP19}.

Instead of mitigating the
effects of memory corruptions, other approaches aim to prevent the
corruptions in the first place by retrofitting C and \Cpp with memory safety, a thorough
enforcement of \emph{temporal} and \emph{spatial} integrity for memory
accesses. Early solutions proposed dialects to the C
language \cite{JMG+02,NMW02}, replacing unsafe language features with
safe alternatives, such as fat pointers. %
However, language dialects impose great restrictions on the
programmer and on the compatibility with existing code, requiring code to be
adapted or rewritten. Later solutions adopted compiler-based or binary-based
instrumentation to equip applications with memory safety, with the
former being the most commonly used form today. Instrumentation-based
approaches refrain from restricting the C and \Cpp languages, and
therefore are completely transparent to the programmer. To achieve temporal memory
safety, solutions typically delay the reuse of freed memory
\cite{Linux-EF,SN05,BZ11,SBPV12,BMCP18}, perform garbage collection \cite{JMG+02,NMW02}, bind pointers to the lifespan of their objects
\cite{ABS94,NZMZ10,SSS+18,vintila2021mesh}, or invalidate pointers on object deallocation
\cite{LSJ+15,You15}. 
To achieve spatial memory safety, solutions either insert
protected memory regions, \eg \emph{red-zones}, between objects \cite{Linux-EF,SN05,BZ11,SBPV12} or track
the bounds of allocated objects or their pointers \cite{ABS94,JMG+02,NMW02,NZMZ09,YPC+10,DY16,DYC17,BMCP18,SSS+18,vintila2021mesh}. However, to provide complete memory safety, solutions must combine temporal
and spatial techniques, usually hindering their practicability %
due to significant performance and memory penalties.

\glsunset{id}
\glsunset{cpu}
In recent years, hardware-supported solutions have started to gain traction.
While memory safety-focused \gls{cpu} extensions, such as CHERI~\cite{WWN+15}
and the ARM \gls{mte}~\cite{armv8.5}, are still actively developed and/or not yet
available in commodity hardware, the ARM \gls{pa}
extension~\cite{arm-pac} is already widely deployed, primarily
in all newer Apple Silicon processors.
The extension adds new instructions to generate and
verify address signatures, so called \glspl{pac}, which are stored in the non-address bits of pointers.
The key used for signing is managed by higher privileged software, \ie
typically the kernel, and cannot be accessed by the protected application.
Since ARM \gls{pa}'s main use case is not the implementation of full memory
safety, existing approaches use it primarily
to implement \gls{cfi} policies, which is straightforward
for function returns~\cite{arm-pac}, and has been done for stack canaries \cite{PCan},
entire call stacks \cite{LNG+21}, and pointers in general \cite{PARTS}.
PTAuth~\cite{PTAuth} has been the first approach using \gls{pa} for memory
safety but targets temporal vulnerabilities only. PACMem~\cite{li2022pacmem}
promises spatial and temporal safety but restricts the number of allocatable
objects, limiting its usability.

We propose \emph{\Name}, a concept for practical \gls{pa}-based
spatial and temporal memory safety, achieving
strong security guarantees for protected applications on ARMv8.3 platforms.
\Name shadows the unsafely accessed memory of an application,
storing a unique \gls{id} for each memory object, with the \gls{id} covering
the whole shadow area corresponding to the object. Whenever an object is
allocated, the \gls{id} is generated and used to create a \gls{pac} specific
to pointers to this object. 
\Name ensures that the association of a pointer to an object and its \gls{id} 
never changes. When a pointer is used, its \gls{pac} is first checked
using the \gls{id} of the location that it tries to access. If
the pointer has been manipulated to point to another object, the verification fails, since the \gls{id} does not
match its \gls{pac}. When an object is deallocated, its \gls{id} is erased from the shadow.
This ensures that all remaining pointers to the object stop
working, since their \gls{pac} cannot be verified without the now erased
\gls{id}.

\newcommand*\rot{\rotatebox{90}}
\newcommand*{\p}[1]{\includegraphics[width=1.5mm, height=1.5mm]{graphics/circle_#1.pdf}}
\begin{table}[tbp]
	\setlength{\tabcolsep}{3pt}
	\small
	\centering
	\caption{Comparison of \Name and related solutions.
		\label{tab:comparison}}
	\vspace{1em}
	\begin{tabular}{lllllll}
		\toprule
		\multicolumn{1}{c}{}
		& \rot{\em Spatial safety}
		& \rot{\em Temporal safety}
		& \rot{\em Security}
		& \rot{\em Interoperability}
		& \rot{\em Perf. overhead}
		& \rot{\em Mem. overhead}\\
		\midrule
		ASan~\cite{SBPV12} &
		\p{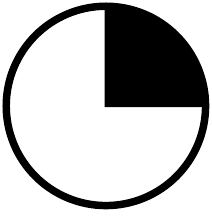} & \p{25} & \p{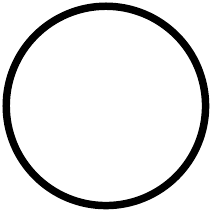} & \p{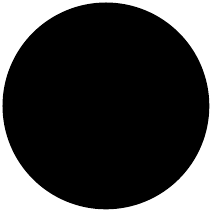} & \p{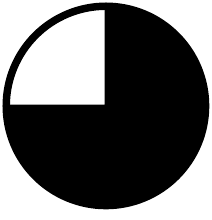} & \p{0}\\
		HWAsan~\cite{SSS+18} &
		\p{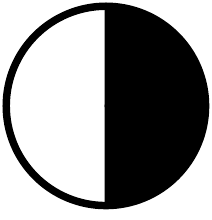} & \p{75} & \p{25} & \p{100} & \p{50} & \p{100}\\ 
		PTAuth~\cite{PTAuth} &
		\p{0} & \p{50} & \p{100} & \p{100} & $\p{100}^*$ & $\p{100}^*$\\
		Softbound\slash CETS~\cite{NZMZ09,NZMZ10} &
		\p{100} & \p{100} & \p{50} & \p{25} & \p{0} & \p{75} \\
		\midrule
		\Name &
		\p{75} & \p{100} & \p{100} & \p{100} & \p{75} & \p{75}\\
		\bottomrule
		\multicolumn{1}{c}{} \\
	\end{tabular}
	\footnotesize
	\centering
	\vspace{1em}
	\parbox[t]{11em}{
		\raggedright
		\emph{Spatial safety}\\
		\p{25} Only linear overflow detection\\
		\p{50} Probabilistic with low entropy\\
		\p{75} No overflow to padding detection\\
		~\\
		\emph{Interoperability}\\
		\p{25} Can pass pointers to uninstrumented libraries\\
		\p{100} Supports externally allocated objects\\
	}
	\hspace{2em}
	\parbox[t]{12em}{
		\raggedright
		\emph{Temporal safety}\\
		\p{25} No protection for pointers to reused memory\\
		\p{50} Only heap protection\\
		\p{75} Probabilistic with low entropy\\~\\~\\
		\emph{Security}\\
		\p{0} No pointer metadata\\
		\p{25} Pointer metadata with low entropy\\
		\p{50} Pointer metadata\\
		\p{100} Cryptographic pointer metadata\\
	}
	\hspace{1em}
	\parbox[t]{7em}{
		\raggedright
		\emph{Perf. overhead}\\
		\p{0} Over 300\%\\
		\p{50} Up to 200\%\\
		\p{75} Up to 150\%\\
		\p{100} Up to 50\%\\
		$\p{100}^*$ ~From \cite{PTAuth}\\~\\~\\
		\emph{Mem. overhead}\\
		\p{0} Over 1000\%\\
		\p{25} Up to 1000\%\\
		\p{75} Up to 300\%\\
		\p{100} Up to 100\%\\
		$\p{100}^*$ ~From \cite{PTAuth}\\
	}
\end{table}

\Name achieves complete temporal and spatial
memory safety, protecting heap, stack, and global objects alike. 
Table~\ref{tab:comparison} compares \Name to highly relevant and/or related solutions
for commodity hardware.
\Name achieves similar or better performance and memory consumption, and an excellent level of memory safety, 
only compromising in the spatial detection of intra-object violations and very small overflows/overreads
into padding bytes.
Additionally, \Name offers high security against crafted pointers due to the
cryptographic properties of \gls{pa}. Its interoperability with external,
uninstrumented libraries makes it very usable in real-world applications.
\gls{asan}~\cite{SBPV12}, with its memory safety based on reuse delays and red-zones,
cannot detect malicious accesses that target reused memory. Also
\emph{non-linear} reads and writes, \ie accesses through a pointer that is not just incremented but set to a new value, can skip red-zones and remain undetected. Furthermore, without
pointer metadata, \gls{asan} offers weak protection against pointer crafting. 
\gls{hwasan} \cite{SSS+18} implements memory tagging for ARM platforms.
With its 8-bit tags, its detection is probabilistic with a comparably high
chance for collisions. While its memory consumption is very low, it performs
considerably worse than \Name.
PTAuth \cite{PTAuth} implements cryptographically secure temporal memory safety but only
protects the heap and does not provide spatial memory safety altogether.
Softbound\slash CETS~\cite{NZMZ09,NZMZ10} provides uncompromising memory safety but
suffers from poor performance and interoperability issues.  \Name combines the
main strengths of the other approaches into a single solution for commodity
hardware, \ie ARM platforms. 
\Name can be used for testing, \eg replacing ASan in fuzzing setups,
but is also a viable mitigation tool protecting itself against attacks
involving pointer crafting.

In summary, we make the following contributions:
\begin{itemize}
	\item A concept for utilizing \gls{pa} to protect the spatial and temporal integrity of memory objects.
	\item \Name, our solution that applies this concept to protect heap, stack, and global objects of C and
	\Cpp applications.
	\item A full LLVM-based prototype implementation\footnote{Source code available under \url{https://github.com/Fraunhofer-AISEC/cryptsan}.} 
	of \Name.
	\item A detailed evaluation of \Name and related solutions showing \Name's
	effectiveness and efficiency on a \gls{pa}-enabled Apple M1
	processor.
\end{itemize}

\section{ARM Pointer Authentication}\label{sec:arm_pac}

\glsreset{pa}
ARM \gls{pa} is a hardware feature in the
ARMv8.3 architecture~\cite{arm-pac,armv8.5}. The feature enables software to
cryptographically protect the integrity of pointers, and hence ensure that
they are not maliciously modified.

The feature introduces keys, managed by a privileged component, \eg the kernel,
that ensures each process is provided a unique key and that keys are changed
correctly upon context switches.
Using these keys, two types of new instructions are provided by the
architecture for signing and authenticating pointers.
The \texttt{pac} instruction \emph{signs} a
pointer and stores the signature, called
the \gls{pac}, in the non-address bits of the pointer. 
This is possible, since those bits are architecturally unused:
If no architectural tagging is used, all non-address bits except bit 55, which is used to
determine if an address is in the \emph{high} or \emph{low} region, are
used for storing the \gls{pac}. Therefore, the size of the \gls{pac} depends
on the configurable size of virtual addresses $n$. With virtual address sizes
between 52 and 32 bits, a \gls{pac} can have between 11 and 31
bits~\cite{arm-pac}. 
On our macOS M1 test system, $n$ is configured to 47 bit, resulting in
16-bit \glspl{pac}.
In addition to the pointer and key, the \texttt{pac} instruction uses a
\emph{context} value of 64 bits, which can be used similarly to a salt value. 

The \texttt{aut} instruction \emph{verifies}
signed pointers, \ie checks their \gls{pac} value, again using the key and context value. 
If key and context are equal to the ones used for
creating the \gls{pac}, and if the pointer has not been modified in the meantime, the verification succeeds and the pointer is made
usable by stripping its \gls{pac} bits. Otherwise, the processor replaces the
\gls{pac} bits with a special bitmask that triggers an exception into the managing
component as soon as the pointer is dereferenced in a load, store, or branch instruction.

\section{\Name Object Protection}\label{sec:pac_object_life_cycle}

To protect the spatial and temporal integrity of objects with \Name, we introduce cryptographically secured pointers.
Specifically, we leverage ARM \gls{pa} to lock pointers to their objects by adding a \gls{pac} in their non-address bits.
Using this \gls{pac}, we are able to guarantee that a pointer may only be used for accessing its pointed-to object during the lifetime of that object.
In this section, we describe the changes to an object's life cycle, namely its allocation, usage, and deallocation, and introduce the metadata required to facilitate the locking.
For our concept, we assume the metadata is kept in a 1:1 \emph{shadow memory}: the available memory is split in halves and every byte is shadowed by one byte of metadata.

\paragraph{Allocation.}
\begin{figure}[t]
	\centering
	\includegraphics[width=\linewidth]{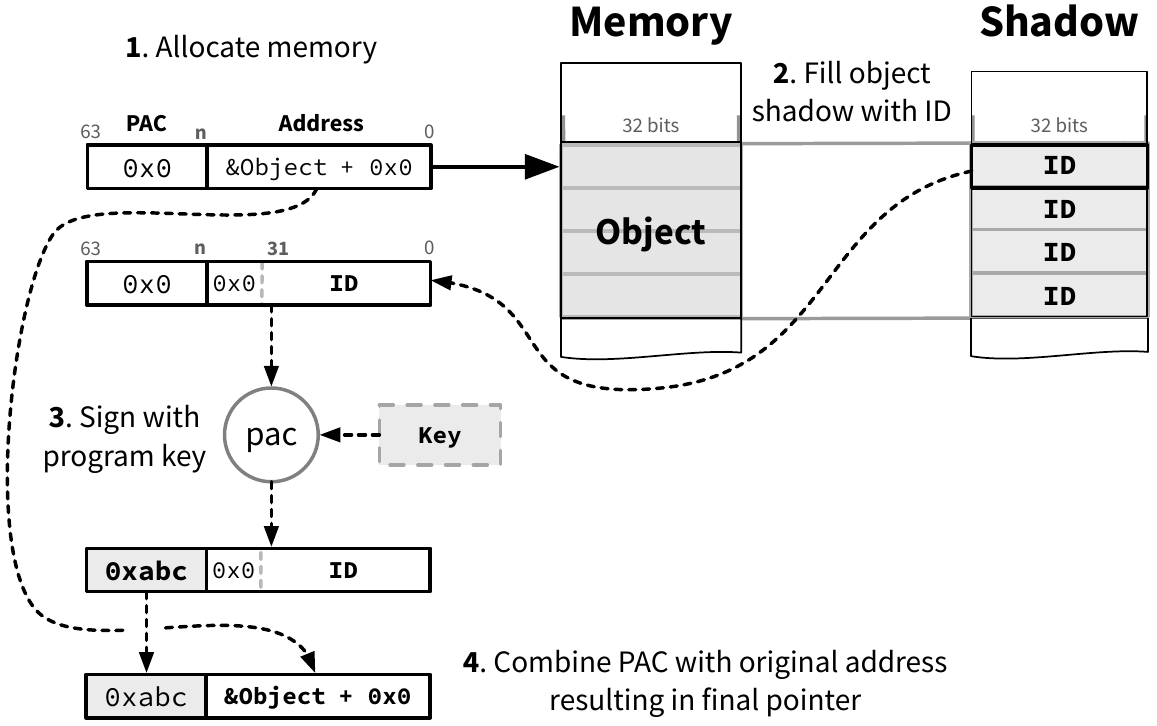}
	\caption{\label{fig:allocation}Allocation of a \gls{pa}-protected object.}
\end{figure}
The first step of the object life cycle is the
allocation. The allocation of an object protected with \gls{pa} is shown in
Figure~\ref{fig:allocation}. As in Section~\ref{sec:arm_pac}, $n$ denotes the size of virtual addresses on
a particular system. The program first allocates memory for the object as it
would normally do, but then also generates a unique 32-bit \emph{identifier} (ID) that
is repeatedly written into the corresponding shadow memory. The ID 
is generated from a global counter, which is incremented after each
allocation and initialized to a random value at program start.
After writing the ID to the shadow memory, the
ID is extended with zero-bits and signed using the \texttt{pac} instruction with a zero context.  
Note that the \gls{msb} of the address bits is \emph{fixed} to
zero, protecting the shadow memory by never signing pointers to it, as
discussed in Section~\ref{sec:memory_layout}.
As discussed in Section~\ref{sec:arm_pac}, the \texttt{pac} instruction calculates a
cryptographic signature, the \gls{pac}, and stores it in the upper bits.
The \gls{pac} is then combined with the original address to create the
final cryptographically secured pointer used for the rest of the object's lifetime.

The ID is crucial for protecting objects with \gls{pa}, as it enables the detection of spatial and temporal memory safety violations by
marking the dimension and liveliness of objects in the shadow memory.
As the ID is repeatedly written into the entire shadow memory covered by newly
allocated objects, the memory actually allocated for objects must be a
multiple of 32 bit. Therefore, memory allocations are padded to fulfill this requirement.
Further, using a 32-bit wide ID ensures that
the input for the signing instruction is always larger than the resulting
\gls{pac}, which is important in regard to certain attacks discussed in
Section~\ref{sec:security}. 

\paragraph{Usage.}

\begin{figure}[t]
	\centering
	\includegraphics[width=\linewidth]{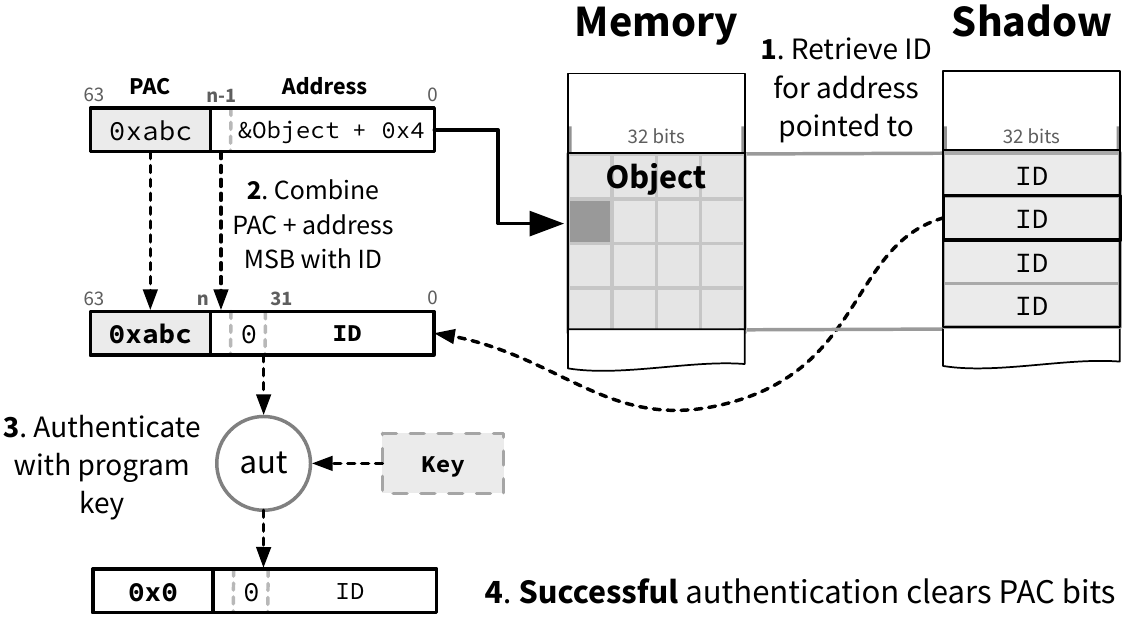}
	\caption{\label{fig:usage_legal}Authentication of a \gls{pa}-protected object.}
\end{figure}
After an object has been allocated, it can be used through its secured pointer
or any pointer derived from it. For \gls{pa}-protected objects, any
pointer to the object carries a \gls{pac} in its non-address
bits. Using this \gls{pac}, it is possible to verify the integrity of the pointer
before allowing access to the object. An example of a legal access to
an offset of 4 byte into a \gls{pa}-protected object is shown in
Figure~\ref{fig:usage_legal}. First, using the aligned pointer address, the ID
corresponding to the pointed-to object is retrieved from the shadow memory. To
authenticate the pointer, its address is then substituted with the 32-bit wide ID, only preserving the \gls{pac} and the \gls{msb} of the address.\footnote{
	As explained in Section~\ref{sec:arm_pac}, the size of virtual addresses $n$ on
	a particular system can typically be configured between 52 and 32. Since the
	address \gls{msb} must \emph{not} be replaced with the ID, for \Name, $n$ has
	to be between 52 and 33, resulting in a \gls{pac} size between 11 and 30 bits.
}
Next, the result is verified using the \texttt{aut}
instruction, reversing the signing operation from the allocation. 
As our example depicts a legal access, the
verification succeeds, clearing the \gls{pac} bits in the process.
Since the address \gls{msb} is always set to zero during the \gls{pac} creation, the verification fails for pointers addressing the upper half of memory where \Name's metadata is stored.
Finally, after ensuring the successful verification, a version of the original
pointer stripped of its \gls{pac} is generated for the actual access.

Apart from being used for accessing memory, pointers are also used for deriving
other pointers, typically through copying or pointer arithmetic.
For signed pointers, copying or arithmetic operations
implicitly carry the pointer's \gls{pac} to the derived pointer,
automatically locking it to the object it is derived from.
Hence, pointer derivation can be left unmodified, and derived pointers can still be verified
as described above and shown in Figure~\ref{fig:usage_legal}.

\begin{figure}[t]
	\centering
	\includegraphics[width=\linewidth]{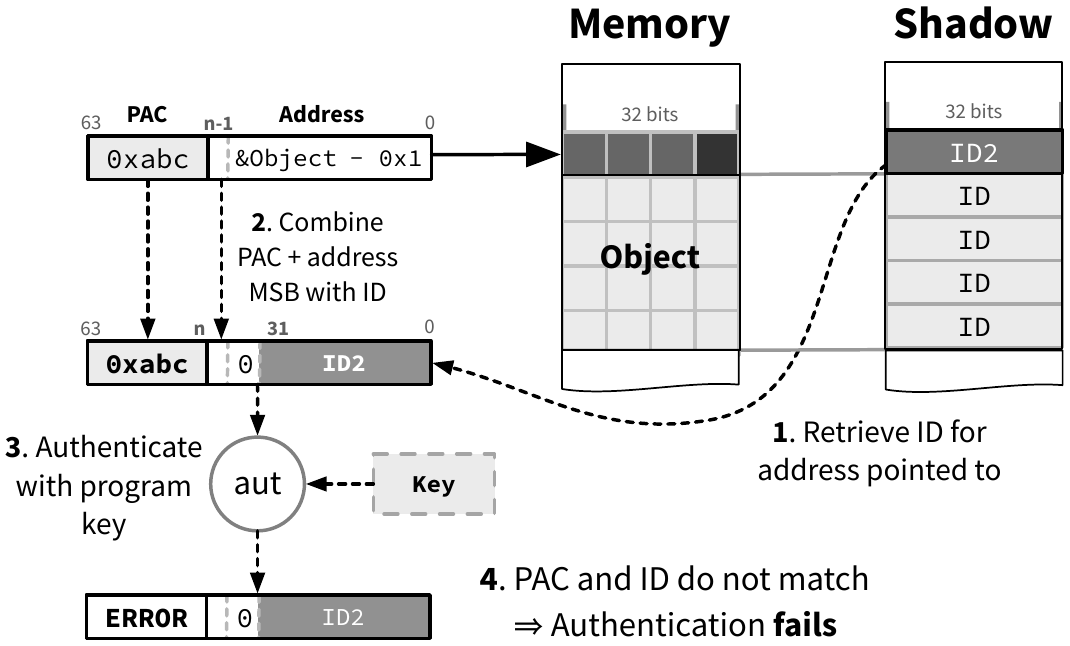}
	\caption{\label{fig:usage_illegal}Underflow on a \gls{pa}-protected object
		as an example for a spatial memory safety violation.}
\end{figure}
Typically, pointer derivation is prone to creating illegal pointers due to faulty
arithmetic operations increasing or decreasing offsets beyond the object's
bounds or overflowing the address into the pointer's \gls{pac} bits. For
illegal pointers whose \gls{pac} bits have been corrupted, the verification
trivially fails.
For illegal out-of-bounds pointers, a verification example 
is shown in Figure~\ref{fig:usage_illegal}. 
Again, the verification process itself is identical
to verifying any other signed pointer.
Using the address the pointer points to, the ID is retrieved from the shadow memory
and the pointer's address bits (except the \gls{msb}) are substituted with it.
However, as the pointer illegally points to a different object, a
mismatching ID is retrieved and the \texttt{aut} instruction fails.
In contrast to a successful authentication where the \gls{pac} bits of the pointer are
cleared, the unsuccessful \texttt{aut} instruction sets the \gls{pac} to an \texttt{ERROR}
bit combination, enabling the detection of the illegal access.

\paragraph{Deallocation.}

\begin{figure}[t]
	\centering
	\includegraphics[width=\linewidth]{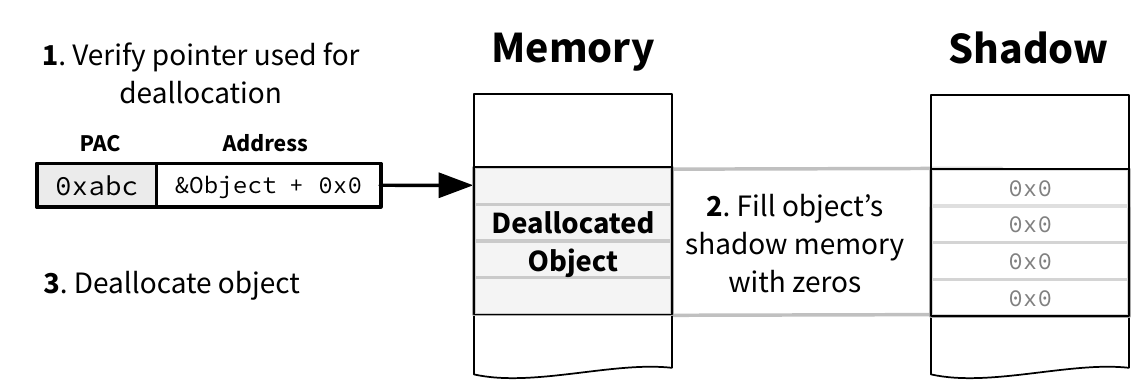}
	\caption{\label{fig:deallocation}Deallocation of a \gls{pa}-protected object.}
\end{figure}

The last step of the object life cycle is the deallocation, in which an object is
destroyed by freeing its allocated memory. 
The deallocation of a
\gls{pa}-protected object is shown in Figure~\ref{fig:deallocation}. Before the object's
pointer can be used for deallocating, its spatial and
temporal integrity is verified, prohibiting a deallocation through an illegal
pointer. For a legal pointer, it is then verified to point to the beginning of
the object to prevent \emph{free-inside-buffer} errors:
The ID corresponding to the pointer and its preceding ID must be unequal, indicating that the pointer points to the \emph{first} ID in shadow memory.
After a successful verification, the object is marked as freed by overwriting its entire shadow memory with zeros.
Finally, the object's memory is freed using the standard deallocation.

Resetting the shadow memory to zero implicitly invalidates all associated pointers, and
hence enables the detection of temporal memory safety violations, such as \emph{double-free} and \emph{use-after-free}.
If the freed memory is reused for a new object later, a new ID is generated and written into the corresponding shadow memory.
The differing IDs of the old and new object ensure that a temporal memory
safety violation on reused memory, namely through a \emph{use-after-free}
error, is prohibited as well, as visualized in
Figure~\ref{fig:usage_illegal_temporal}.

\section{\Name Program Protection}\label{sec:program_protection}

\Name uses per-object metadata to lock pointers to their
objects: objects are assigned unique IDs and pointers to these
objects are signed using those IDs. In this section, we present how \Name
embeds this locking into programs to achieve memory safety.

\subsection{Enforcing Memory Safety}\label{sec:protected_objects}

\begin{figure}[t]
	\centering
	\includegraphics[width=\linewidth]{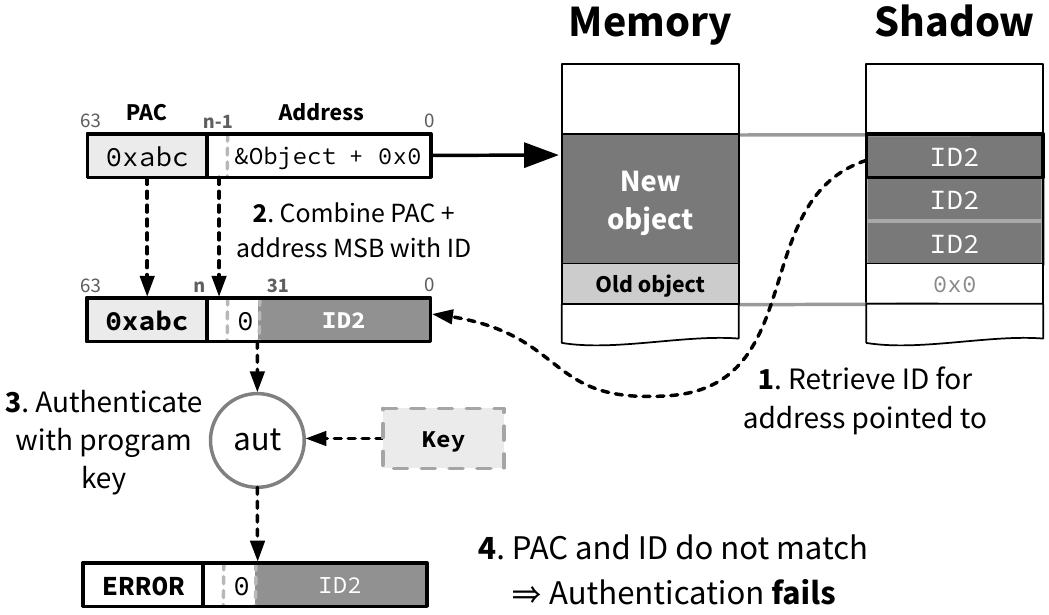}
	\caption{\label{fig:usage_illegal_temporal}Use-after-free on
		reused memory as example for a temporal memory safety
		violation on a \gls{pa}-protected object.}
\end{figure}

With \Name, we protect heap, stack, and global objects.
\Name instruments the allocation and deallocation of these objects to maintain their metadata and lock their pointers with \glspl{pac}.
To validate the locked pointers before allowing memory to be accessed, \Name also instruments all uses of pointers with \emph{checks}.
These checks implement the authentication of \glspl{pac} as described in Section~\ref{sec:pac_object_life_cycle} and shown in Figures~\ref{fig:usage_legal}, \ref{fig:usage_illegal}, and \ref{fig:usage_illegal_temporal}.
However, depending on the object's location, type, and usage,
\Name might omit instrumentations if it is safe to do so and slightly adapts
instrumentation during allocation and deallocation if necessary.
For example, while address-taken objects are always instrumented, directly
accessed local variables are not because they are always accessed safely.
In the following, we discuss the differences in instrumenting heap, stack, and global objects.

\paragraph{Heap Objects.}
In C and \Cpp heap objects are always accessed through pointers.
Thus, \Name always instruments them straightforwardly
as described in
Section~\ref{sec:pac_object_life_cycle}.

\paragraph{Stack Objects.}

Unlike on the heap, the majority of objects on the stack are
local variables that are safely accessed directly through the stack pointer.
Those objects cannot be used for memory corruptions but can themselves be
compromised through corruptible---\ie \emph{unsafe}---objects also stored on the stack.
To protect objects on the stack, it is therefore sufficient to only
instrument unsafe objects, preventing them from compromising each
other or safe objects. To distinguish between safe and unsafe objects,
we use the approach proposed by SafeStack \cite{KSP+14}: %
Stack objects are generally considered safe, unless their addresses are taken
or their accesses cannot statically be guaranteed to be in-bounds.
According to this distinction, \Name instruments unsafe stack objects, generating IDs and \glspl{pac} in function prologues, clearing the IDs in function epilogues, and inserting checks before the objects are accessed.
Clearing the IDs in the epilogue enables \Name to detect \emph{use-after-scope} errors.
For unsafe stack objects whose addresses are \emph{not} taken and which are
therefore accessed through the stack pointer, \Name still constructs signed pointers and redirects accesses through them to instrument checks as usual.
With this instrumentation, \Name guarantees the temporal and spatial integrity of the entire stack, without requiring changes to safe objects or the stack's layout.
Merely the padding to a multiple of 32 bit for instrumented objects extends the memory required by the stack slightly.

\paragraph{Global Objects.}
\glsunset{pic}

For global objects, \Name applies the same static analysis as for the stack,
only instrumenting unsafe objects that might potentially corrupt other objects.
For these objects, their global scope poses an
additional challenge compared to heap and stack objects.
Since the key used for \gls{pa} changes on each program run,
\Name cannot generate signed pointers statically
but must create them at run-time.
Further, as global objects are also allocated statically, non-\gls{pic} executables access them statically as well, requiring no pointers to be used at all.

To secure unsafe global objects, 
we introduce the \gls{gppt}, which contains a signed pointer for each
unsafe global object and is accessible as global variable itself.
During instrumentation, accesses to the original global objects are
modified to first load the corresponding signed pointer from the \gls{gppt} and
then authenticate the pointer to ensure spatial and temporal correctness.
The code for generating the ID,
signing the pointer, and storing the pointer in the
\gls{gppt} is instrumented at the beginning of the main function of the
instrumented program.
Since all accesses to the \gls{gppt} are added by \Name itself, loads and
stores to it are not instrumented with checks.

\subsection{\textbf{Memory Layout and Metadata}}\label{sec:memory_layout}

\Name relies on per-object metadata stored in shadow memory, as presented in Section~\ref{sec:pac_object_life_cycle}.
The shadow must store unique IDs over the full
length of objects.
To realize this, \Name alters the memory layout of protected
programs. In particular, \Name divides the entire virtual address space in
halves, creating a 1:1 mapping between usable memory in the lower half and
corresponding shadow memory in the upper half. 
While the lower half can be used by the program to store code and data, the
shadow memory in the upper half is exclusively managed by \Name to store
metadata for instrumented objects.
With the 1:1 mapping, addresses of instrumented objects can easily be translated to their
corresponding shadow addresses by setting their \gls{msb}.
This enables \Name to perform efficient
metadata lookups. Note that the \gls{msb} of the address is \emph{set}, not flipped, so that the
mapping is a one way operation and addresses in the shadow map to themselves.

Further, also as described in Section~\ref{sec:pac_object_life_cycle}, the address \gls{msb} is forced to zero during allocation.
This guarantees that \Name can only generate signed pointers for objects in
the lower half of the address space and that instrumented checks will only
succeed if the address \gls{msb} of checked pointers is zero.
Hence, pointers trying to access the shadow memory directly by setting their
address \gls{msb} \emph{always} fail authentication.
This is also the case for pointers trying
to access their own metadata, which would, without the incorporation of the address
\gls{msb}, have a \gls{pac} matching the ID in the shadow.
Metadata security is further discussed in
Section~\ref{sec:security}.

\subsection{\textbf{Interoperability}}\label{sec:external}

With \Name, we aim to achieve full interoperability with
uninstrumented code, especially for the use of dynamically linked libraries. 
To support signed pointers being passed to uninstrumented code, \Name detects calls to external functions and instruments the stripping of the \gls{pac} for pointer arguments.
This allows external functions to access instrumented objects as usual without requiring any modifications.
The distinction between internal and external functions is possible by
postponing \Name's instrumentation to the \glsdesc{lto} stage, where truly external
functions can be identified.

For pointers coming from uninstrumented code, \Name must create valid signatures to allow them being used properly by the instrumented code.
For this, \Name instruments external function calls returning pointers to
read the corresponding ID from the shadow and using that ID to generate a valid \gls{pac}.
Note that this assumes that returned objects are always shadowed with an ID.
This is trivially the case for pointers to objects allocated
in instrumented code and passed to an external function. %
To also support pointers to externally allocated objects, \Name is able to intercept
external allocations via the libc allocator, and ensure that an ID is generated and written to each object's shadow memory.
The interception is completely transparent to the library and forwards the
actual allocation of memory to the originally called routine.

\subsection{Optimizations}\label{sec:optimizations}

Principally, \Name has to instrument each memory access to achieve memory
safety. Nonetheless, in several situations checks can be omitted because the
underlying memory access is statically safe, as already explained in
Section~\ref{sec:protected_objects}, or redundant, as detailed in the
following.

\paragraph{Redundant Check Removal.} 
If multiple memory accesses use the exact same pointer, \ie address and type,
and one of the accesses \emph{always} executes before the others, \ie it
\emph{dominates} them, it is sufficient to check only the first access.
\Name performs an intra-procedural analysis to detect and remove such
redundant checks. 
This approach was adapted from Softbound's
optimization for removing redundant spatial checks \cite{NZMZ09}. Since \Name's
checks combine spatial and temporal authentication, our analysis
additionally verifies that between two memory accesses with the same pointer,
the pointer is not passed to functions possibly freeing the underlying memory.

\paragraph{Same Lock Optimization.} 
Consecutive memory accesses read or write to the same object, but increment or
decrement the pointer each time. In \Name, this pointer derivation results in
every pointer containing the same \gls{pac} and every access through these
pointers being legal if the \gls{pac} can be authenticated using the pointed-to
object's ID. However, as soon as \emph{one} of these pointers has been
authenticated successfully, any subsequently derived pointer can trivially be
verified by checking whether it points to an object shadowed by the same
ID. This \emph{fast} verification is possible, as the relationship between
\gls{pac} and ID has already been validated and can only be invalidated by a
changing ID, \ie the derived pointer overflowing into the neighboring object.

To apply this optimization, similar to the redundant check removal, \Name
identifies load and store instructions accessing the same object with different
offsets and then instruments the normal pointer verification for the dominating
instruction while instrumenting the fast verification for all dominated
instructions. Again, the analysis verifies that between the dominating and
dominated instructions the pointer is not passed to a function possibly
freeing the underlying memory. 

\section{Implementation}\label{sec:implementation}

Our \Name prototype is implemented as an extension to the LLVM compiler framework
(version 12) and supports macOS on ARMv8.3 hardware, \ie M1 Macs. Our
LLVM extension provides an instrumentation pass and a custom runtime
library. 
Our instrumentation pass primarily inserts checks before pointer uses and redirects library
calls, \eg for allocation and deallocation.
Additionally, for function calls
to uninstrumented code, the pass instruments the stripping of signed pointer
arguments and the signing of returned pointers.
Instrumented programs are linked against our runtime library, which 
initializes and maintains the shadow memory and
provides wrappers and interceptors for commonly used functions in the C/\Cpp
standard libraries that require checks or allocate memory.
These implementation aspects are discussed in more detail in the following.

\paragraph{Shadow Memory Initialization.} To initialize the shadow memory, the
runtime library uses a constructor, which is executed by the
dynamic linker when loading an instrumented program. This constructor
simply allocates the upper half of the virtual memory using \texttt{mmap}.
If this fails, the runtime library aborts the program.

\paragraph{Shadow Memory Maintenance.}
\Name maintains metadata for heap, stack, and global objects as described in
Sections~\ref{sec:pac_object_life_cycle} and \ref{sec:protected_objects}. For allocations on
the heap, \Name provides
wrappers for the standard C/\Cpp allocation and deallocation routines (\ie
\texttt{malloc}, \texttt{free}, etc.) that implement the additional handling of
metadata. 
Our instrumentation pass replaces calls to the standard routines with
calls to the \Name wrappers, which in turn call the standard routines to do
the actual allocation and deallocation.
For stack and global objects, the generation of metadata is directly instrumented into
the program. Global objects remain allocated for the entire lifetime of the program so
that \Name initializes their metadata only once at load-time.

\paragraph{Standard Library Wrappers.}

When linking instrumented programs against pre-compiled C/\Cpp standard
libraries, library functions taking pointer arguments, such as \texttt{memcpy}, may be used
to access and manipulate instrumented objects. To guarantee memory safety
in those cases, our runtime library
provides wrappers that check and strip signed pointers before forwarding them to the
originally called function. Note that these functions may also return
one of their pointer arguments, for which the wrapper simply
returns the signed pointer as received as argument. Again,
our instrumentation pass replaces
calls to the library functions with calls to their
corresponding wrappers.

\paragraph{Standard Library Interceptors.}

For interoperability with pre-compiled libraries, \Name is able to create
signed pointers even when heap objects are allocated by uninstrumented code.
The runtime library also provides interceptors for the standard C/\Cpp allocation and deallocation routines that manage IDs even for heap objects allocated in uninstrumented code.
In contrast to the internal allocation wrappers,
the interceptors return unsigned pointers, directly usable by the
uninstrumented code. Pointers are then re-signed in the instrumented code after function return.

\section{Security Discussion}\label{sec:security}

For our discussion, we assume an uncompromised system running programs
that contain memory vulnerabilities. An attacker may target those
vulnerabilities with the intent to leak program-internal data or corrupt the
programs' control flow. However, the attacker cannot interfere with programs
through the system, \eg by manipulating the PA keys in the kernel.
Within these attacker capabilities, \Name is designed to prevent any malicious
manipulation of pointers, as discussed in the following. 

\paragraph{Illegally Derived and Dangling Pointers.}
An attacker might compromise a pointer without corrupting the
\gls{pac} itself (see the next paragraph for attacks that also corrupt the
\gls{pac}). For example, they might exploit faulty pointer arithmetic to
manipulate a pointer's address bits or use an unmodified but dangling pointer. 
As described in Sections~\ref{sec:pac_object_life_cycle} and \ref{sec:program_protection},
\Name identifies potentially unsafely accessed objects and 
locks pointers to their object's lifetime and bounds.
Illegal accesses via pointers not pointing to their original object only
succeed if a \emph{signature collision} occurs,
\ie if the \gls{pac} happens to be the same for both objects. This is
trivially the case if two objects share the same ID, a \emph{metadata 
	collision}, but due to the limited \gls{pac} size, can also happen
for objects with different IDs. Depending on the virtual address
size a system is using, \Name's \gls{pac} size $p$ is between 11 and 30 bits (tagging
disabled). On macOS the \gls{pac} is 16 bits wide. The probability for 
a signature collision when targeting a specific object is $\frac{1}{2^p}$, \ie 
only $\frac{1}{2^{16}} = 1.52\cdot10^{-5}$ for \Name on macOS.
Furthermore, since these collisions happen unpredictably, we consider their
exploitation a low risk.
\Name's 32-bit wide IDs ensure that the full \gls{pac} space is used for all
possible \gls{pac} sizes. 
Metadata collisions might be predictable if the randomly initialized
counter for ID generation is exhausted in a single program execution. But since that
only happens after $2^{32}-1$ allocations, %
we consider this an insignificant risk.

\paragraph{Crafted Pointers.}
Typical vulnerabilities only allow an attacker to compromise the address bits
of a pointer or to use a dangling pointer. Nonetheless, under certain conditions,
an attacker might be able to completely \emph{craft} a pointer including
its \gls{pac} bits. This can, for example, happen via
very large pointer arithmetic overflows into the \gls{pac} bits or via
\emph{intra-object} corruptions, in which the targeted pointer is inside a
compound object together with an unsafe member, \eg an array. 
In such a case,
\Name's cryptographic pointer protection ensures that an attacker crafting a
pointer must guess the right \gls{pac} for the object they want to access,
with a low probability of $\frac{1}{2^{p}}$ to succeed, as described
before. As discussed in the next paragraph, this is also the case even if the
attacker is able to leak the ID of the targeted object from the shadow memory.

\paragraph{Metadata Security.}
In contrast to other approaches based on pointer tagging, such as HWASan (see~Section~\ref{sec:related_work}), the
cryptographic nature of \glspl{pac} ensures that metadata access does not
directly compromise the security of corresponding objects.
The relationship between metadata and \gls{pac} is protected by the key, which
is managed by a privileged component and inaccessible to the program.
An attacker who is able to leak the ID of an object and craft a pointer, is still not able to generate a
valid signature without knowledge of the key.
Even being able to alter the ID of an object does not enable an attacker to generate a
valid signature.
Additionally, as described in Section~\ref{sec:memory_layout}, \Name prevents direct access to metadata.

\paragraph{Uninstrumented Libraries.} \Name allows protected programs to
link against pre-compiled libraries. 
When linked against uninstrumented
code, \Name still provides its full protection inside the
instrumented code. 
Even if objects are allocated by uninstrumented code, \Name generates metadata and signs the pointer once it is returned to instrumented code.
However, an attacker may illegally access protected objects 
through vulnerabilities within the uninstrumented code, where memory accesses are
not checked.

\section{Evaluation}\label{sec:evaluation}

We evaluated our \Name prototype on an M1 MacBook
Pro 2020 (16 GB memory) with actual hardware support for ARM
\gls{pa} running macOS Big Sur in version 11.3.1.
In our evaluation, we tested whether \Name reliably detects
temporal and spatial memory corruptions, and how much performance and memory
overhead \Name induces at run-time. To measure \Name's memory safety
guarantees we
used the Juliet Test Suite\footnote{\url{https://samate.nist.gov/SRD/testsuite.php}}. To measure the performance and
memory overhead we used the SPEC CPU
2017\footnote{\url{https://www.spec.org/}} benchmark suite.
We performed the evaluation in comparison to memory safety solutions related to \Name, namely
ASan~\cite{SBPV12}, \gls{hwasan}~\cite{SSS+18}, and Softbound\slash
CETS~\cite{NZMZ09,NZMZ10}. 
We could not measure the overheads of PTAuth since the provided setup only simulates ARM \gls{pac} on x86\_64 using dummy instructions.

\subsection{Vulnerability Detection Rate}
\label{sec:evaluation_precision}

\begin{table}[tb]
	\small
	\centering
	\caption{Juliet Test Suite corruption prevention ratio.}
	\label{table:juliet}
	\begin{tabular}{l r r r r}
		\toprule
		\textbf{CWE} & \textbf{\Name} & \textbf{ASan} & \textbf{HWASan} & \textbf{Softb./CETS} \\
		(\# cases) & (5364) & (5364) & (5364) & (3970) \\
		\midrule
		121: Stack Overflow 				& 98.5\% 	& 96.7\% 	& 82.9\% 	& 77.7\% \\
		122: Heap Overflow					& 97.4\% 	& 94.7\% 	& 94.6\% 	& 73.7\% \\
		124: Buf. Underwr. 					& 100\% 	& 100\% 	& 81.9\% 	& 82.5\% \\
		126: Buf. Overrd. 					& 100\% 	& 100\% 	& 99.7\% 	& 96.5\% \\
		127: Buf. Underrd. 					& 100\% 	& 100\% 	& 75.9\% 	& 78.4\% \\
		415: Double Free 					& 100\% 	& 100\% 	& 100\% 	& 100\% \\
		416: Use After Free 				& 100\% 	& 83.0\% 	& 50.9\% 	& 51.3\% \\
		761: Free Inside Buf. 				& 100\% 	& 100\% 	& 0\% 		& 100\% \\
		\midrule
		Overall								& 98.7\%	& 97.0\%	& 85.4\%	& 78.9\%	\\
		\bottomrule
	\end{tabular}
\end{table}

The Juliet Test Suite is a collection of C/\Cpp test
cases covering a variety of \glspl{cwe} including temporal
and spatial memory corruptions. For each test case there is a \emph{good}
version without any memory violations and a \emph{bad} version containing a
memory corruption. 
The suite also provides several variants, wrapping the same
vulnerability in different control flow or data flow constructs, \eg embedding the vulnerable object in a structure.
For our evaluation, we excluded variants that depend on external input (\eg
\texttt{fgets}) to easily enable automation. 
We also removed variants with control flows that only occasionally or never
exhibit an actual memory violation at run-time, resulting in a total of 5364 test cases. 
For the evaluation of Softbound\slash CETS, additional cases leading to compilation
errors had to be excluded, resulting in a total of 3970 cases.

Table~\ref{table:juliet} shows the results of evaluating \Name
and related solutions with Juliet. None of the approaches showed false positives, \ie
crashes in the good variants of the test cases. 
\Name mitigates all temporal and spatial bugs, except intra-object overflows (34 test cases).
Note that \Name does not detect violations if an overflow within the object padding (see Section~\ref{sec:pac_object_life_cycle}) occurs,
which is the case for some of the stack and heap overflow tests.
Still, these overflows are successfully mitigated as they do not compromise the security of the program.
Compared to the other solutions, \Name achieves a higher or equal mitigation rate for all CWEs.
Like \Name, none of the other solutions detect intra-object overflows, although Softbound\slash CETS should conceptually.
ASan further misses some wrapper functions resulting in a slightly lower mitigation rate than \Name.
For HWAsan, missing wrappers and an incomplete temporal bug detection account for most of the additional non-mitigated bugs.
While Softbound\slash CETS should have the best accuracy conceptionally, the released source code lacks features and also misses various wrapper functions.

Taking \Name's protection against pointer crafting into account, it would further achieve 100\% mitigation rates for the stack and heap overflow tests, as the intra-object test cases overflow into a pointer corrupting the \gls{pac}, which \Name detects before the pointer is dereferenced.
Furthermore, important corruption categories that would 
further differentiate \Name from ASan, most notably
non-linear overflows, are not tested by Juliet at all. 

\subsection{Performance and Memory Overhead}

To measure \Name's overhead,
we used the SPEC CPU 2017 benchmark suite, comparing the
performance and memory consumption of benchmarks compiled with \Name and
the other solutions to a baseline of the benchmarks compiled without. 

We compiled all benchmarks with optimization level \texttt{O2} and executed them single-threaded.
Out of the $17$ C/\Cpp \emph{rate}
benchmarks, we had to exclude \textit{502.gcc}, because of a memory violation
detected by all mechanisms. Additionally, we had to exclude \textit{511.povray},
\textit{523.xalancbmk}, and \textit{526.blender} due to known
incompatibilities with \gls{pa} on macOS \cite{applecpp}: These
benchmarks contain virtual methods whose declarations differ from their definitions
causing them to crash on execution when compiled with LLVM 12
and \gls{pa}.
We further had to port Softbound\slash CETS from LLVM 3.9 to a recent LLVM version to compile and run SPEC CPU 2017.
Our ported version works with most benchmarks, but still fails to compile 5 out of 13 benchmarks due to errors in the instrumentation.
We also measured Softbound\slash CETS on x86, as its the only supported architecture.

\begin{table}[tb]	
	\small
	\centering
	\caption{\label{tab:overheads}Measured performance overhead of memory safety frameworks for SPEC CPU 2017 benchmarks.}
	\begin{tabular}{l r r r r}
		\toprule
		\textbf{Benchmark} & \textbf{\Name} & \textbf{ASan} & \textbf{HWASan}$^*$ & \textbf{Softb./CETS}$^{**}$ \\
		\midrule
		500.perlbench   & $92.8 \%$ & $113.6\%$	& $347.0  \%$  & - 	\\
		505.mcf         & $80.2 \%$ & $45.4 \%$	& $96.5 \%$  & $548.9 \%$\\
		508.namd        & $139.3\%$ & $69.3 \%$	& $386.5\%$  & $350.8 \%$\\
		510.parest      & $109.0\%$ & $79.8 \%$	& $300.3\%$  & -	\\
		519.lbm         & $28.8 \%$ & $11.6 \%$	& $44.5 \%$  & $148.0 \%$\\
		520.omnetpp     & $96.9 \%$ & $107.9\%$ & $130.7\%$  & -	\\
		525.x264        & $114.0\%$ & $166.4\%$ & $98.7 \%$  & $419.4 \%$\\
		531.deepsjeng   & $170.4\%$ & $80.7 \%$	& $135.4\%$  & $510.7 \%$\\
		538.imagick     & $155.0\%$ & $143.9\%$ & $275.8\%$  & $609.7 \%$\\
		541.leela       & $76.4 \%$ & $66.3 \%$	& $122.8\%$  & -	\\
		544.nab         & $67.6 \%$ & $126.6\%$ & $104.8\%$  & $243.4 \%$\\
		557.xz          & $55.1 \%$ & $40.3 \%$	& $76.4 \%$  & -	\\
		999.specrand    & $203.5\%$ & $380.2\%$ & $58.9 \%$  & $46.1 \%$	\\
		\midrule
		Average 		& $106.8\%$ 	& $110.2\%$ & $167.5 \%$  & $359.6 \%$\\
		Geo. Mean 		& $95.5\%$ 				& $83.0\%$ 	& $134.7\%$ 	& $283.5\%$ 	\\
		\bottomrule
	\end{tabular}
	\parbox[t]{\linewidth}{
		\small
		\vspace{2mm}
		$^*$~No support for macOS. Therefore evaluated with Linux on Apple M1. \\
		$^{**}$~No support for ARM architectures. Therefore evaluated on x86-64. 
	}
\end{table}
\subsubsection{Performance}
\label{sec:evaluation_performance}
The results of our performance evaluation are shown in
Table~\ref{tab:overheads}.
The evaluation shows that \Name achieves an average performance overhead of
$106.8\%$ which is comparable to ASan and better than the other solutions. %
Looking at single benchmarks, the overhead varies from small to
substantial. We attribute this difference to the varying memory usage, with
some benchmarks iterating over large amounts of memory requiring
\Name to verify memory accesses frequently and resulting in a significant
performance overhead. Other benchmarks mainly
process local, safely accessed variables on the stack, resulting in fewer
memory accesses verified by \Name and a comparably small performance overhead.

\subsubsection{Memory Overhead}
By maintaining object IDs, \Name imposes a memory overhead on protected
programs. Each \gls{pa}-protected object is shadowed 1:1, causing 100\% overhead.
While this relationship is clear, there are several aspects that make an
estimation of a program's actual overhead much less straightforward, justifying an
experimental evaluation. 
First, objects are padded to 32-bit boundaries, possibly increasing
the overhead beyond 100\%.
Second, as explained in Section~\ref{sec:protected_objects},
not all objects require explicit protection for the program to be memory safe:
For the stack and globals, \Name only protects and shadows \emph{unsafe}
variables, generally reducing the overall overhead. Third, freeing a
shadowed object does \emph{not} free its
shadow but overwrites it with zero.
While we expect the allocator to eventually reuse much of the
memory, which also reuses its shadow, this behavior leads
to overheads of more than $100\%$ for most programs.

To evaluate \Name's memory overhead, we
compared the peak memory consumption of the instrumented benchmarks,
using \texttt{time -l} on macOS and \texttt{time -f \%M} on
Linux, %
against the corresponding non-instrumented baselines. The results are summarized in
Table~\ref{table:mem_overhead}. \Name achieves an average memory overhead of $193.4\%$.
As expected, the overhead heavily depends on the actual workload, ranging from
$6\%$ for the \emph{999.specrand} benchmark, not using dynamically allocated
memory, to $680.9\%$ for the heap-intensive \emph{500.perlbench} benchmark.

\begin{table}[tb]	
	\small
	\centering	
	\caption{Measured memory overhead of memory safety frameworks for SPEC CPU 2017 benchmarks.}
	\label{table:mem_overhead}
	\begin{tabular}{l r r r r}
		\toprule
		\textbf{Benchmark} & \textbf{\Name} & \textbf{ASan} & \textbf{HWASan}$^*$ & \textbf{Softb./CETS}$^{**}$ \\
		\midrule
		500.perlbench   & $680.9 \%$ & $311.0  \%$ & $34.3 \%$& -         \\
		505.mcf         & $124.0 \%$ & $55.1   \%$ & $8.9 \%$& $638.0\%$ \\
		508.namd        & $110.7 \%$ & $222.9  \%$ & $20.6 \%$& $88.3 \%$ \\
		510.parest      & $259.6 \%$ & $894.2  \%$ & $54.7 \%$& -         \\
		519.lbm         & $99.8  \%$ & $15.6   \%$ & $6.4 \%$& $1.9  \%$ \\
		520.omnetpp     & $94.8  \%$ & $380.7  \%$ & $27.9 \%$& -         \\
		525.x264        & $119.5 \%$ & $110.2  \%$ & $4.7 \%$& $261.4\%$ \\
		531.deepsjeng   & $98.5  \%$ & $14.3   \%$ & $20.3 \%$& $2.2  \%$ \\
		538.imagick     & $180.2 \%$ & $170.5  \%$ & $9.6 \%$& $106.5\%$ \\
		541.leela       & $472.7 \%$ & $11558.2\%$ & $28.3 \%$& -         \\
		544.nab         & $134.8 \%$ & $156.6  \%$ & $62.3 \%$& $40.7 \%$ \\
		557.xz          & $133.0 \%$ & $44.8   \%$ & $4.5 \%$& -         \\
		999.specrand    & $6.0   \%$ & $2445.6 \%$ & $43.6 \%$& $390.9\%$ \\
		\midrule
		Average 	& $193.4 \%$ & $1260.0 \%$ & $25.1 \%$ & $191.2 \%$ \\
		Geo. Mean 	& $128.0 \%$ & $204.3 \%$ & $17.9 \%$ & $56.5 \%$\\
		\bottomrule
	\end{tabular}
	\parbox[t]{\linewidth}{
		\small
		\vspace{2mm}
		$^*$~No support for macOS. Therefore evaluated with Linux on Apple M1. \\
		$^{**}$~No support for ARM architectures. Therefore evaluated on x86-64. 
	}
\end{table}

\subsubsection{Comparison}
\Name performs similar to or better than the related solutions with the
only exception being the lower memory overhead of \hwasan.
Compared to ASan, \Name achieves similar performance and undercuts the memory overhead by
almost a magnitude, while offering significantly better spatial and temporal memory
safety. Compared to Softbound\slash CETS, \Name achieves much improved performance
while retaining the same level of memory consumption and memory safety, with
its only compromise being the loss of spatial \emph{detection} precision (not
security) for overflows (of less than 4 bytes)
into \Name's padding. Compared to \hwasan which uses a 16-to-1 mapping between application memory and shadow metadata, \Name imposes higher memory
overhead but considerably better performance and memory safety guarantees.
Technical details of the tested solutions and other related approaches are
discussed in the next section (see also Table~\ref{tab:comparison} for a comparison
summary).

\section{Related Work}\label{sec:related_work}
Protecting C and \Cpp applications against attackers trying to corrupt memory has been an active field of research for the past decades.
Performance-oriented solutions focus on code pointers only, protecting return addresses on the stack with canaries or enforcing \gls{cfi} \cite{BCN+17,BZP19}.
PCan \cite{PCan}, PACStack \cite{LNG+21}, PARTS \cite{PARTS}, and PACTight
\cite{281318} utilize \gls{pa} to cryptographically protect stack canaries,
entire call stacks, and pointers to code and/or data, respectively. 
In contrast to \Name and memory safety approaches in general, these solutions aim to
detect corrupted pointers instead of preventing an attacker from corrupting them in the first place.
As such, they can be implemented very efficiently but
do not protect all application data against memory bugs, such as out-of-bounds
or use-after-free accesses.

The least precise memory safety solutions protect objects by interleaving them
with \emph{guard pages} \cite{Linux-EF,DA06b} or \emph{red-zones}
\cite{HJ91,SN05,BZ11,HMS12,SBPV12}.
These solutions, including the popular sanitizer ASan~\cite{SBPV12}, are able
to prevent \emph{linear} overflows between objects, but cannot detect
accesses skipping guard pages or red-zones.
To approximate temporal memory safety, they commonly \emph{delay} or
\emph{prohibit} the reuse of allocated memory at the cost of possible memory exhaustion.
\Name's approach provides superior precision, preventing any spatial and temporal corruption between memory objects.

Other solutions maintain \emph{per-object} metadata to guarantee spatial memory safety only \cite{JK97,RL04,DA06a,ACCH09,YPC+10,DY16,DYC17}.
They store bounds information of objects in a disjoint data structure and
enforce pointer arithmetic to stay within the original object, with the
exception of temporary, non-dereferenced out-of-bounds pointers, which must be
handled explicitly.
Since pointers lack metadata, these solutions are not able to detect temporal memory corruptions.
For spatial memory safety, they provide almost optimal precision, only leaving intra-object corruptions 
and overflows into potential padding undetectable.
\Name also maintains per-object metadata, but locks pointers to their
pointed-to objects using non-address pointer bits.
This eliminates the need to handle out-of-bounds pointers explicitly and enables \Name to provide temporal memory safety.

Similar to our work, \hwasan \cite{SSS+18} extends the idea of maintaining metadata per object by utilizing unused pointer bits to store a link to the pointed-to object's metadata.
In \hwasan, the per-object and per-pointer metadata are identical: For each object, an 8-bit identifier is randomly chosen and stored in the unused bits of pointers to that object.
Simultaneously, the identifier is kept in a memory-aligned shadow memory, effectively providing a \emph{lock} to the pointers' \emph{keys} in their unused bits.
This allows \hwasan to provide temporal memory safety without maintaining individual metadata per pointer.
Before dereferencing a pointer, \hwasan verifies that the pointer's key is identical to the lock stored in the shadow memory corresponding to the pointed-to object.
However, with its identifiers of only 8 bits in size, \hwasan is prone to
metadata collisions leading to undetected memory corruptions, as shown in our evaluation where it only detects 75.9\% of the buffer underreads in Juliet. 
Furthermore, for \hwasan, metadata security is critical, whereas \Name's
cryptographic approach can tolerate, for example, metadata leaks.
Additionally, in \Name, the hardware supported check logic leads to a significant performance increase compared to \hwasan.

Developed at the same time as \Name, PACMem \cite{li2022pacmem} uses the \gls{pac} instructions as hashing functions to efficiently access an object's bounds in a metadata table.
This design increases performance but also limits the amount of allocated
objects based on the \gls{pac} length. 
On macOS with 16 bits wide \glspl{pac}, this would allow programs to allocate
only $2^{16}$ objects, which is quite limiting, considering that this includes
heap \emph{and} stack allocations.

Approaches with \emph{per-pointer} metadata typically achieve the highest precision.
First solutions utilized fat pointers to store bounds information for spatial safety checks \cite{Ste92,ABS94,JMG+02,NMW02} and typically deployed garbage collection for temporal safety \cite{JMG+02,NMW02}.
Later solutions, such as Softbound\slash CETS \cite{NZMZ09,NZMZ10}, opted for
keeping bounds information disjoint in memory and used  \emph{lock and key} metadata to perform temporal safety checks \cite{PF97,XDS04,NZMZ09,NZMZ10}.
However, maintaining metadata per pointer degrades run-time performance
significantly and yields poor interoperability with uninstrumented code.
Compared to these solutions, \Name maintains per-object metadata but utilizes unused pointer bits to lock pointers to their pointed-to objects.
With this hybrid approach, \Name has comparable protection against spatial and
temporal memory corruptions, while achieving better run-time performance and
providing better interoperability with uninstrumented code.

Some solutions focus on detecting temporal memory corruptions only, either by \emph{tracking pointers} on a per-object basis \cite{CGMN12,LSJ+15,You15} or, similar to \Name, by \emph{locking pointers} to their pointed-to objects \cite{NZMZ10,PTAuth}.
Tracking all pointers pointing to an object is straightforward, as only allocation and pointer derivation must be monitored.
Whenever an object is deallocated, all registered pointers are immediately invalidated.
However, this simple approach cannot be combined or extended with any kind of spatial memory corruption detection, making it inferior to locking-based solutions such as \Name.
PTAuth \cite{PTAuth} utilizes \gls{pa} to sign heap pointers with a 64-bit
identifier prepended to their pointed-to heap objects.
In contrast, \Name provides both spatial and temporal memory safety for heap, stack, and global objects alike.

Finally, with CHERI~\cite{WWN+15} and ARM \gls{mte}~\cite{armv8.5} memory
safety solutions with dedicated hardware extensions have started to gain
traction in the recent years.
However, these are still in development and/or not yet
available on commodity hardware, while \Name can already be deployed on all newer Apple Silicon processors.

\section{Conclusion}
We proposed \Name, a new memory safety concept for C/\Cpp programs, leveraging ARM \gls{pa}.
\Name is able to provide full memory safety, protecting a program's heap,
stack, and globals against both spatial and temporal memory corruptions. 
Since \Name's protection is based on signatures using secret keys,
it provides additional security against attacks on metadata compared to
similar, software-only approaches.
We implemented a fully functional, LLVM-based \Name prototype for macOS on M1 Macs.
In our evaluation
the \Name prototype 
outperforms similar software-based solutions. 
This, together with its robust interoperability with uninstrumented
libraries, makes \Name a viable solution for retrofitting memory safety to C/\Cpp programs.

\begin{acks}
	This work was supported by the \grantsponsor{FhG}{Fraunhofer Internal
		Programs}{} under Grant No. \grantnum{FhG}{PREPARE 840 231}.
\end{acks}

\bibliographystyle{ACM-Reference-Format}

\bibliography{bibliography/bib-short.bib}

\end{document}